\documentclass[11pt,twoside]{article}
\usepackage{asp2010}

\resetcounters

\allowtitlefootnote

\begin{document}

\title{Spectroscopic Observations of
  \boldmath$\delta$\unboldmath\,Sco Through the 2011 Periastron
  Passage\footnotemark}
\author{Th.~Rivinius,$^{1}$ S.~\v{S}tefl,$^1$ D.~Baade,$^{2}$
  A.~C.~Carciofi,$^{3}$ S.~Otero,$^4$ A.~S.~Miroshnichenko,$^5$ and N.~Manset$^6$
%
%
\affil{$^{1}$ESO---European Organisation for Astronomical Research in the Southern Hemisphere, Chile}
\affil{$^{2}$ESO---European Organisation for Astronomical Research in the Southern Hemisphere, Germany}
\affil{$^{3}$Instituto de Astronomia, Geof{\'\i}sica e Ci{\^e}ncias
Atmosf{\'e}ricas, Universidade de S\~ao Paulo, Brazil}
\affil{$^{4}$Buenos Aires, Argentina \& American Association of Variable Star
Observers, Cambridge, USA}
\affil{$^{5}$ Dept. of Physics and Astronomy,
University of North Carolina at Greensboro, USA}
\affil{$^{6}$CFHT, Hawai'i, USA}
}
\titlefootnote{Partly based on observations obtained at the
  Canada-France-Hawaii Telescope (CFHT) which is operated by the National
  Research Council of Canada, the Institut National des Sciences de l'Univers
  of the Centre National de la Recherche Scientifique of France, and the
  University of Hawaii, and on observations collected at the European
  Organisation for Astronomical Research in the Southern Hemisphere, Chile
  under Prog-IDs 077.D-0605 and 087.A-9005.}
\begin{abstract}We present prelimiary results from a coordinated spectroscopic
campaign in 2011, centered on the ${\delta}$\,Sco periastron passage in
July. Data have mostly been obtained with the FEROS/2.2\,m at La Silla and
ESPaDOnS/CFHT at Mauna Kea echelle instruments. Main results include the
absence of tidally induced disturbance to the main $\beta$\,Cephei pulsation
mode and the absence of tidally triggered mass-ejection at time of periastron
proper. The observed (as far as yet analyzed) variations are compatible with
the picture of a disk that is disturbed on its outer radius, with the
disturbance propagating inwards {\it after} the periastron.
\end{abstract}

\begin{figure}[t]\centering
\includegraphics[width=0.6\textwidth]{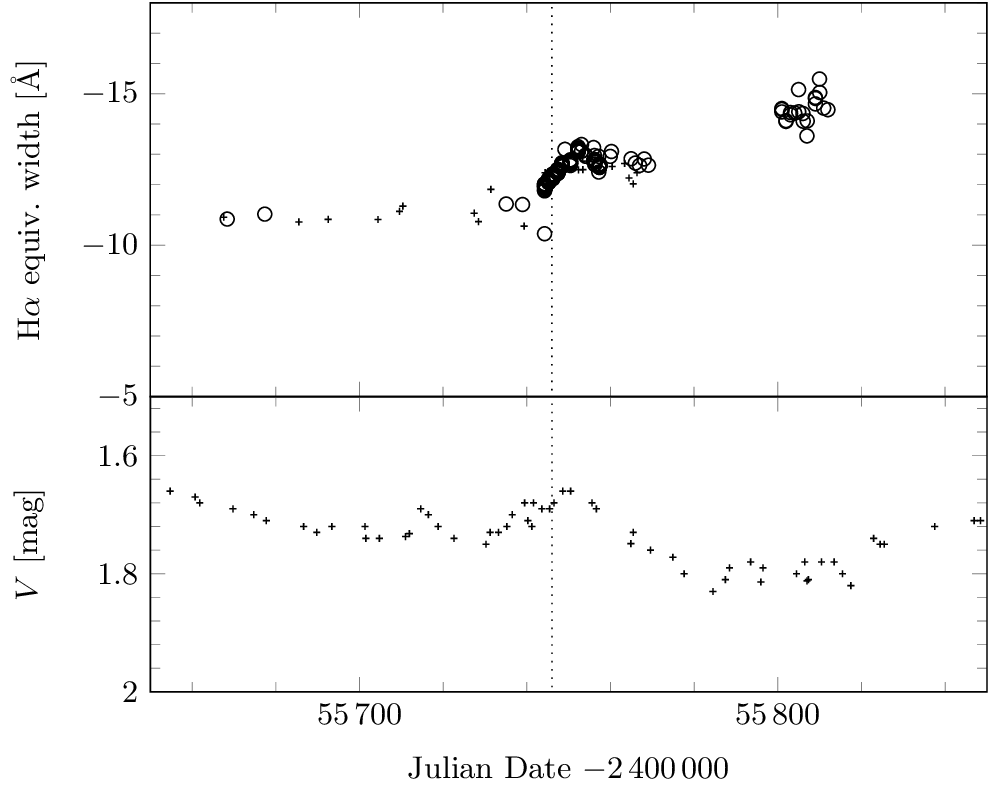}
\caption{\label{fig:ew_peri} The H$\alpha$ equivalent width and $V$-band
  photometry over the periastron (marked by a dotted line). Circles represent
  own spectroscopic data, plus-signs are from the BeSS
  database \citep{2011AJ....142..149N}. The visual photometry is by SO
  (Details will be in Rivnius et al.\, in prep.).}
\end{figure}

\section{Introduction}
The bright B0\,IV $\beta$\,Cephei type star $\delta$\,Sco (7\,Sco, HR\,5953,
HD\,143275) has drawn considerable interest in the past decade (see, for
instance, contributions by Miroshnichenko et al., Jones et al., \v{S}tefl et
al., and Bednarski et al.\ in these proceedings). It is a {high eccentricity
binary} ($e=0.94)$ with a period of about 10 years. In July 2000, close to a
periastron, a rapid brightening occurred, and spectra showed H$\alpha$
emission, making it the second brightest Be star. Initially thought to be an
example of {tidal mass ejection during periastron}, it became clear afterwards
that the periastron occurred about two months after the brightening, in
September, and the disk was present already months before (see Miroshnichenko
et al., these proceedings for a timeline and references). In anticipation of
the next periastron (July 4, 2011, MJD\,55\,746), the object has been observed
intensely.  Here we show the results from spectroscopic observations in 2011.

\begin{figure}[t]\centering
\includegraphics[width=0.4\textwidth]{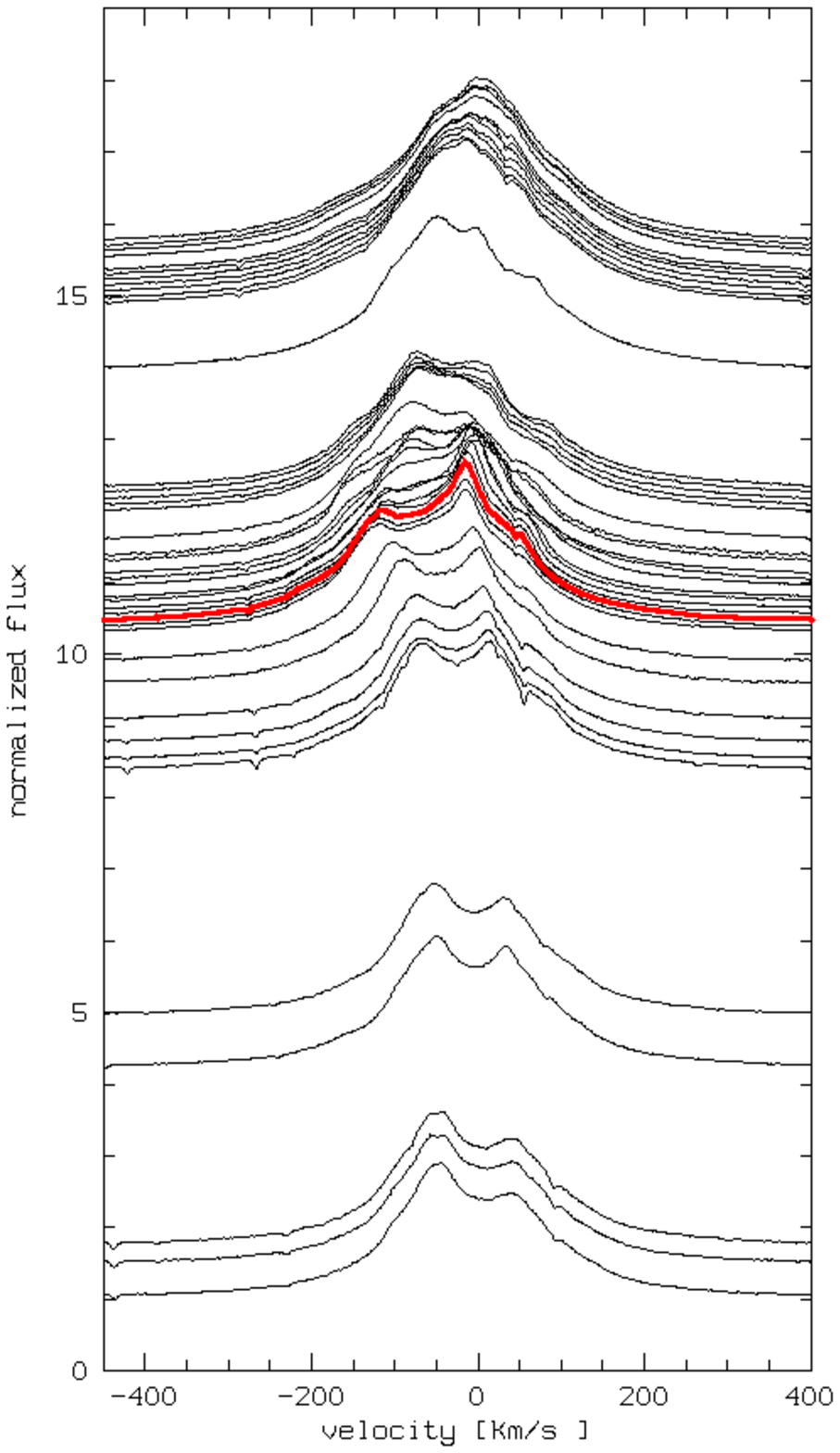}%
\includegraphics[width=0.4\textwidth]{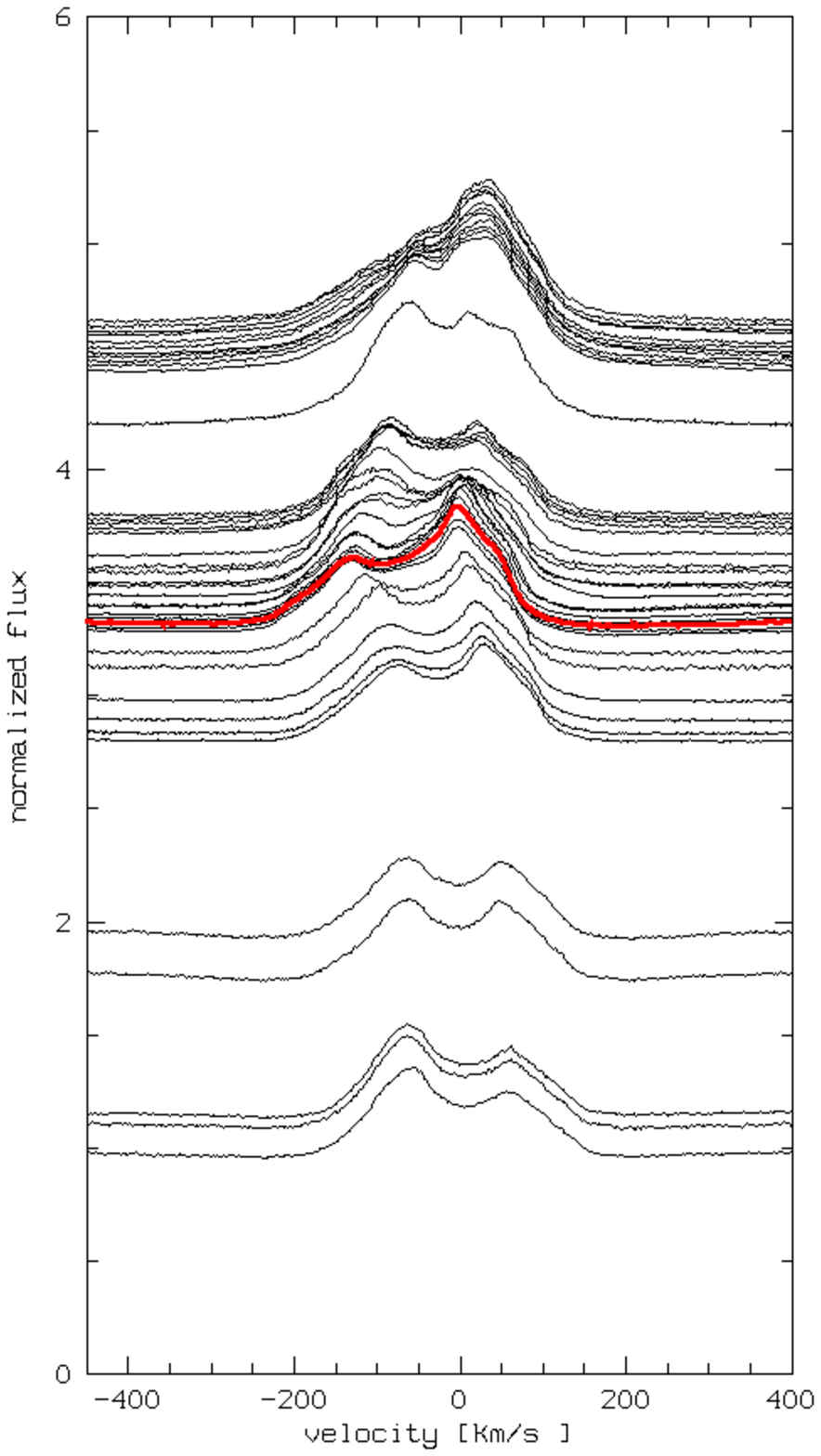}

\caption{\label{fig:HabTS}The H$\alpha$ ({\it left}) and H$\beta$ ({\it
    right}) profile variations in 2011. The vertical offset is proportional to
  time, increasing upwards. The profile at periastron is marked in red.}
\end{figure}

\section{Observations}
Observations were carried out with FEROS at the 2.2\,m telescope on La Silla and
with ESPaDOnS at the CFHT on Mauna Kea. Both are echelle instruments, with
$R=48\,000$ and $68\,000$, respectively, and a coverage from about 370 to
880\,nm. $S/N$ was well above 100 in all observations.  In 2011, a total of
{50 FEROS spectra} were taken in 32 nights between Jan.\ 24 and Sep.\ 8,
while {ESPaDOnS took 304 individual spectra} in 14 nights between Jun.\ 8
and Aug.\ 17. Figure~\ref{fig:ew_peri} shows the evolution of the $V$-band
magnitude together with the H$\alpha$ equivalent width in the year 2011, with
the periastron being marked by a dotted line.

\begin{figure}[t]
\includegraphics[viewport=55 193 326 450,clip,angle=0,width=1.66in]{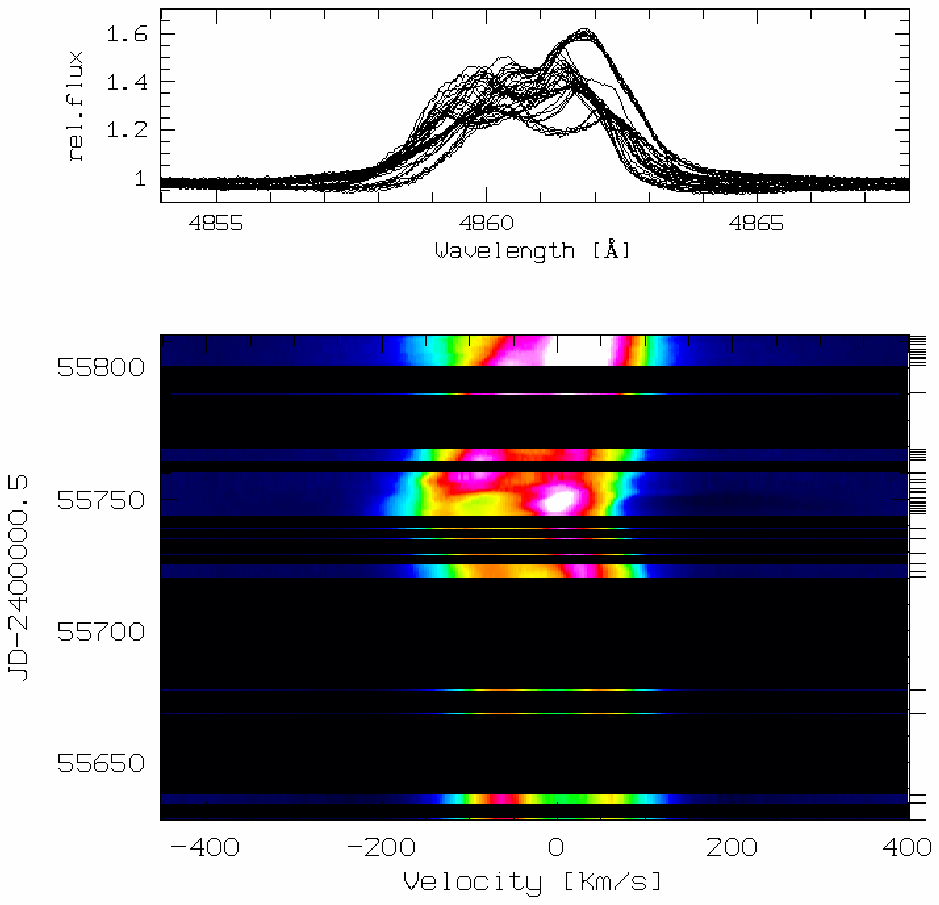}%
\includegraphics[viewport=55 193 326 450,clip,angle=0,width=1.66in]{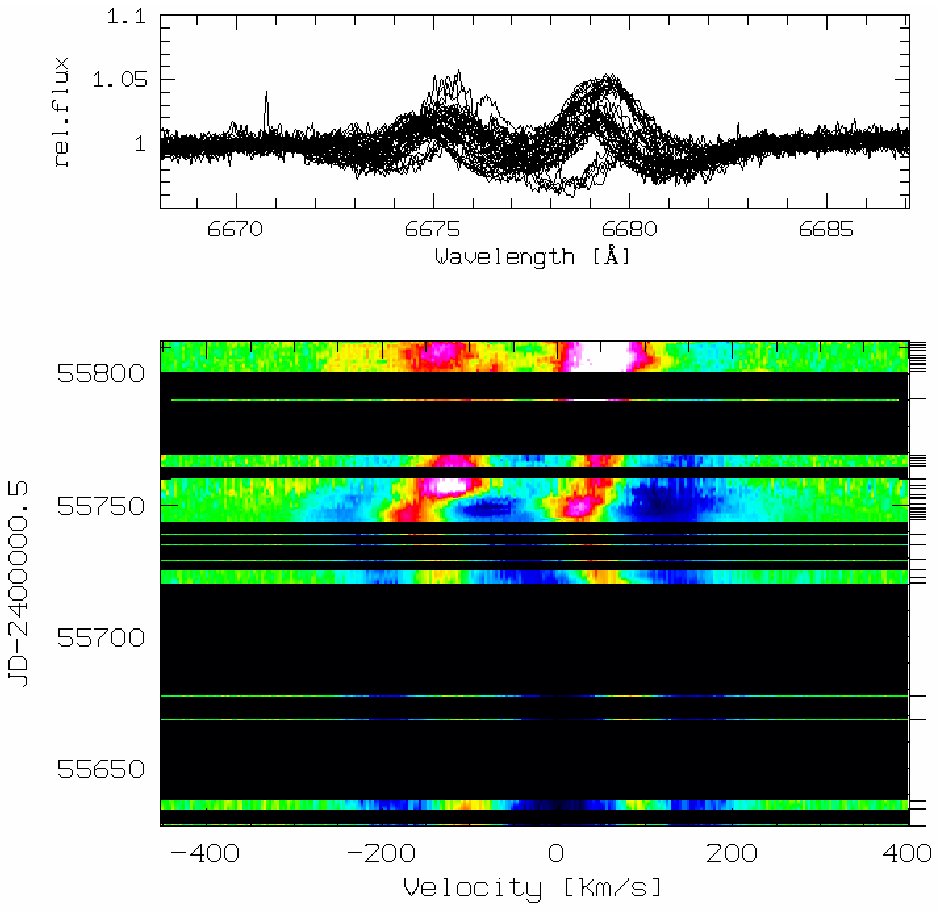}%
\includegraphics[viewport=55 193 326 450,clip,angle=0,width=1.66in]{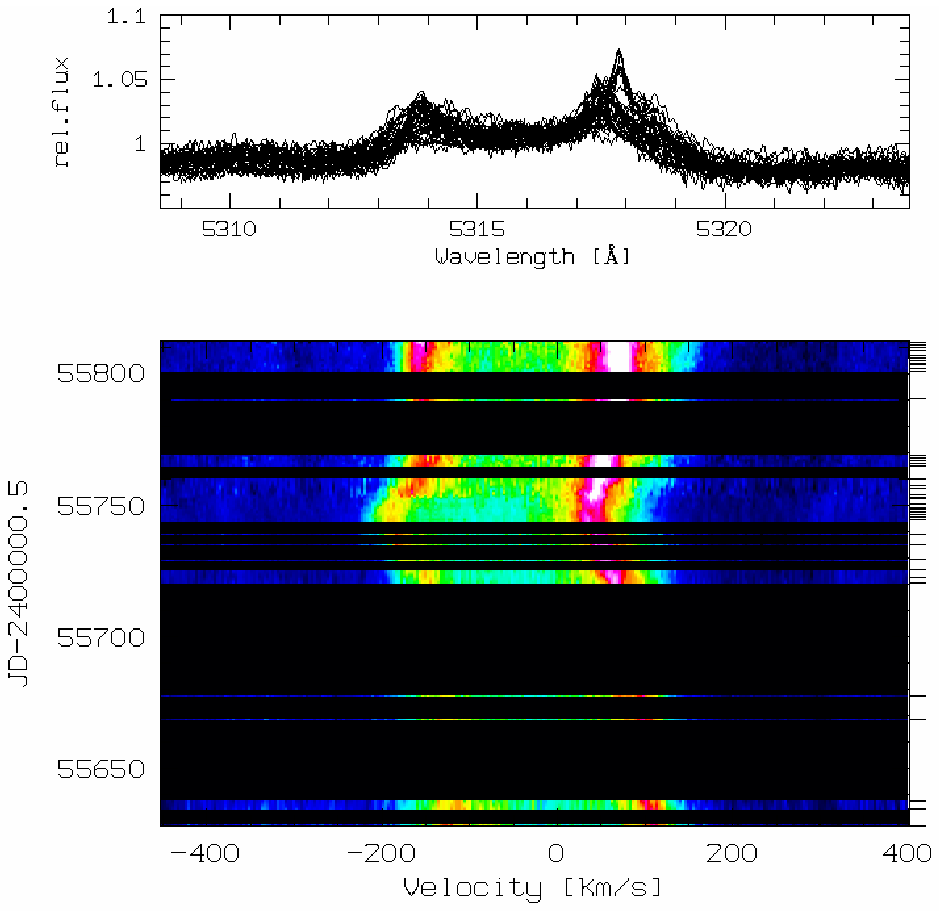}%

\caption[xx]{\label{fig:dynspec_dsc}Line profile variations of emission lines
  observed in 2011. From left to right: H$\beta$, He{\sc i}\,6678, and Fe{\sc
    ii}\,5317. The periastron at MJD\,54\,746 coincides with quadrature.}
\end{figure}

\section{Emission Line Variations}
\emph{Before periastron} (the choice of ``before'' MJD\,55\,401--55\,413, ``during'' 55\,744--55\,748, and ``after''
    55\,760--55\,811 is governed by the availability of spectroscopic data,
    details will be given in Rivinius et al, in prep.), the equivalent width
    of H$\alpha$ {was well stable} at about $-11$\,\AA, while the brightness
    was slowly decreasing (see {Fig.\ref{fig:ew_peri}}).  As close as one week
    to periastron, the emission showed {Be-star typical $V/R$ asymmetry}. The
    value of $V/R$ was slowly decreasing. $V=R$ was reached in H$\beta$ a few
    weeks before periastron (see Fig.~\ref{fig:HabTS}).

\emph{During the periastron}, in which the primary approaches and the
secondary recedes from the observer, excess emission was seen in H$\alpha$ and
H$\beta$, on the red side (see Fig.~\ref{fig:HabTS} and
Fig.~\ref{fig:dynspec_dsc}, left). It remained there for several weeks after the
periastron as an extended ``shoulder''. If orbital and disk angular momentum
vector are aligned (see, however, \v{S}tefl et al., these proceedigns), this
is {in the region between primary and secondary}. The inner disk, probed by
spectral lines of e.g.\ {He}\,{\sc i} and Fe\,{\sc ii} did not show such an
excess.

He\,{\sc i} lines probe a smaller region, i.e. closer to te primary. During
periastron, these lines showed a {moderate increase in emission}, and only
some weeks \emph{after periastron} the emission started to fill in the wings
at high velocity (see Fig.~\ref{fig:dynspec_dsc}, middle).

Finally, Fe\,{\sc ii}\,5317 and similar metal emission lines showed the
  {largest peak separation}, i.e.\ are formed {closest to the primary}. Their
  profiles remained {remarkably undisturbed}, just slightly increasing in
  strength (see Fig.~\ref{fig:dynspec_dsc}, right).

\section{Stellar Pulsation}
\begin{figure}[t]
\plotone{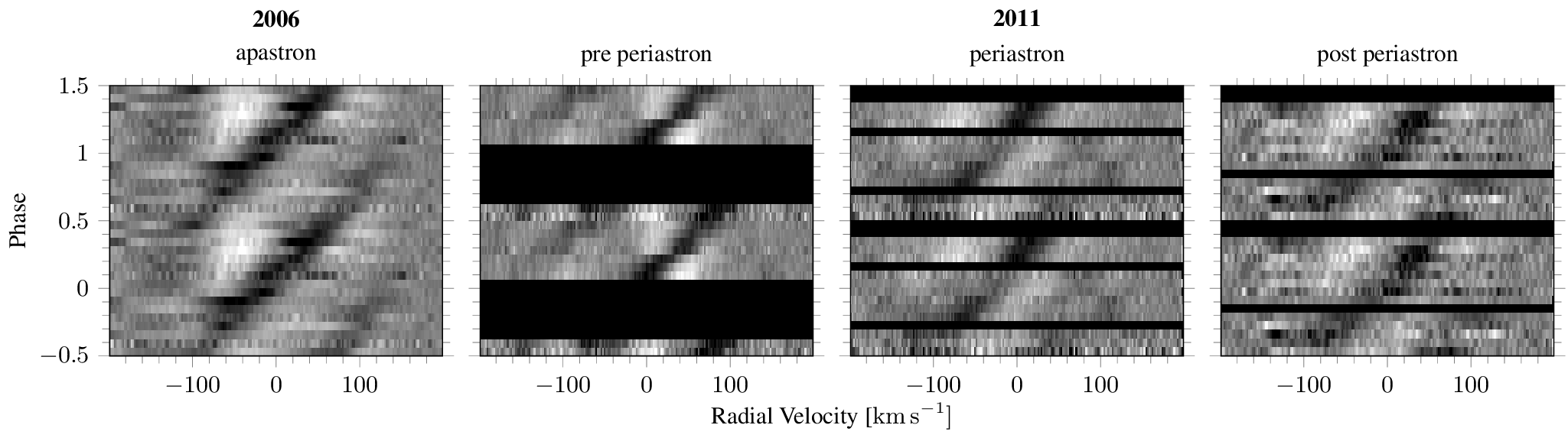}
\caption{\label{fig:bcep}The pulsational $\beta$\,Cephei \emph{lpv} of the
  residuals vs.\ the mean of the \ion{Si}{iii}\,4553 line in 2008
  (MJD\,53\,889--54\,007), and 2011 before (55\,401--55\,413), during
  (55\,744--55\,748), and after (55\,760--55\,811) periastron. Note
  that two full cycles are shown.}
\end{figure}

\citet{1986ApJ...304..728S} reported line profile variability (\emph{lpv}) in
$\delta$\,Sco, being interpreted as non-radial pulsation.  The period is about
2.3\,h (0.097589(3)\,d), with a typical $\beta$\,Cephei appearance. Before,
throughout, and after the periastron, {period, phasing, amplitude, and
  pattern} of the \emph{lpv} {was unaltered}, apart from a small phase shift
fully consistent with the light travel time effect (see
Fig.~\ref{fig:bcep}). {Tidal forces had no impact} on the main $\beta$\,Cephei
pulsation mode of $\delta$\,Sco~A.

\section{Conclusions}
\begin{itemize}
\item {Before the periastron}, the circumstellar
environment of the Be star $\delta$\,Sco\,A {was not affected} by the
approaching  companion.
\item {One week before periastron} on July 4,
H$\alpha$ equivalent width and photometric variability started, peaking
with a strong {hydrogen excess emission} a few days after periastron.
\item The {closer to the primary} an emission line is
formed, {the less and the later} it is affected by {tidal
  perturbations}.
\item The {pulsational behavior} of $\delta$\,Sco A did not change at all
  through the periastron, {no tidal effect} was observed.
\end{itemize}

\bibliography{csdyn}
\end{document}